\documentclass[%
 reprint,
 amsmath,amssymb,
 aps, superscriptaddress,
]{revtex4-1}

\usepackage{graphicx}
\usepackage{dcolumn}
\usepackage{bm}
\usepackage{setspace}
\usepackage{xcolor}
\usepackage{titlesec}
\let\origtau\tau 
\renewcommand{\tau}{\scalebox{1.25}{$\origtau$}}
\usepackage[mathlines]{lineno}
\titlespacing\section{0pt}{12pt plus 4pt minus 2pt}{3pt plus 1pt minus 1pt}
\titlespacing\subsection{0pt}{12pt plus 4pt minus 2pt}{3pt plus 1pt minus 1pt}

\begin{document}

\preprint{APS/123-QED}

\title{Probing Stark-induced nonlinear phase variation with opto-optical modulation}

\author{E.~R.~Simpson}
\email{emma.simpson@fysik.lth.se}
\affiliation{Department of Physics, Lund University, P.O. Box 118, SE-221 00 Lund, Sweden.}%

\author{M.~Labeye}
\affiliation{Louisiana State University, Baton Rouge, 70803-4001, Louisiana, United States of America.}%

\author{S.~Camp}
\affiliation{Louisiana State University, Baton Rouge, 70803-4001, Louisiana, United States of America.}%

\author{N.~Ibrakovic}%
\affiliation{Department of Physics, Lund University, P.O. Box 118, SE-221 00 Lund, Sweden.}%

\author{S.~Bengtsson}%
\affiliation{Department of Physics, Lund University, P.O. Box 118, SE-221 00 Lund, Sweden.}%

\author{A.~Olofsson}%
\affiliation{Department of Physics, Lund University, P.O. Box 118, SE-221 00 Lund, Sweden.}%
 
\author{K.~J.~Schafer}
\affiliation{Louisiana State University, Baton Rouge, 70803-4001, Louisiana, United States of America.}%

\author{M.~B.~Gaarde}
\affiliation{Louisiana State University, Baton Rouge, 70803-4001, Louisiana, United States of America.}%

\author{J.~Mauritsson}%
\email{johan.mauritsson@fysik.lth.se}
\affiliation{Department of Physics, Lund University, P.O. Box 118, SE-221 00 Lund, Sweden.}%

\date{\today}

\begin{abstract}
We extend the recently developed technique of opto-optical modulation (OOM) to probe state-resolved ac-Stark-induced phase variations of a coherently excited ensemble of helium atoms. In a joint experimental and theoretical study, we find that the spatial redirection of the resonant emission from the OOM process is different for the low-lying $1s2p$ state as compared with the higher-lying Rydberg states, and that this redirection can be controlled through the spatial characteristics of the infrared (IR) probe beam. In particular, we observe that the intensity dependence of the IR-induced Stark phase on the $1s2p$ emission is nonlinear, and that the phase accumulation changes sign for moderate intensities. Our results suggest that OOM, combined with precise experimental shaping of the probe beam, could allow future measurements of Stark-induced phase shifts of excited states. 

\end{abstract}

\pacs{Valid PACS appear here}

\maketitle
\section {INTRODUCTION}
Light-matter interactions can be addressed from two complementary points of view \cite{calegari_advances_2016}. Just as light can be used as a tool to probe and control matter \cite{wu_theory_2016,cao_noncollinear_2016,cao_near-resonant_2016,baker_probing_2006, chini_sub-cycle_2013, shivaram_attosecond-resolved_2012}, atoms can be exploited to probe and control light \cite{bengtsson_spacetime_2017, bengtsson_ultrafast_2019, drescher_extreme-ultraviolet_2018, Kaldun_PRL_2014, Blattermann_OL_2015, fleischer_spin_2014, kfir_generation_2015}. The recently demonstrated technique of opto-optical modulation (OOM) \cite{bengtsson_spacetime_2017,bengtsson_ultrafast_2019} is an example of this duality in the realm of ultrafast extreme ultraviolet (XUV) sources. OOM relies on the combination of two coherent femtosecond pulses with different properties. First, an XUV pump pulse resonantly excites an atomic target producing a coherent superposition of ground and excited states. This triggers a long-lived emission of coherent XUV light at  the resonant transition frequencies. 
Subsequently, a strong, infrared (IR) probe pulse arrives and modifies the XUV emission,  altering its spatiotemporal profile. 

The effect of the IR probe pulse on the coherent XUV emission is mediated by the ac-Stark shift \cite{delone_ac_1999}. This IR-induced shift of the excited state energies yields an additional state-dependent {\it phase} that is imprinted on the dipole and thus on the emitted XUV light \cite{wu_theory_2016,ott_lorentz_2013,mashiko_characterizing_2014}. The OOM technique translates the spatial intensity variation of the IR beam into a state-specific spatial phase gradient that results in the redirection of the XUV emission.

Previously OOM has been used to redirect ultrafast XUV light pulses in an argon gas, from both Rydberg and autoionizing states \cite{bengtsson_spacetime_2017,bengtsson_ultrafast_2019}. Further details of the technique using also helium and neon gases can be found in Ref. \cite{bengtsson_ultrafast_2019}.
The direction of emission in these experiments was explained via the known, approximately linear ac-Stark shift of high lying Rydberg states. For these states the ac-Stark shift approaches the average kinetic energy of a free electron oscillating in an electric field, namely the ponderomotive energy $U_\mathrm{p}=e^2F^2/4m_e\omega^2$, where $e$ and $m_e$ are the electron charge and mass, and $F$ and $\omega$ are the electric field amplitude and angular frequency. The ponderomotive Stark phase depends linearly on the IR intensity, which acts as a control parameter on the XUV spatiotemporal properties.

\begin{figure}[t]
\includegraphics[width=\linewidth]{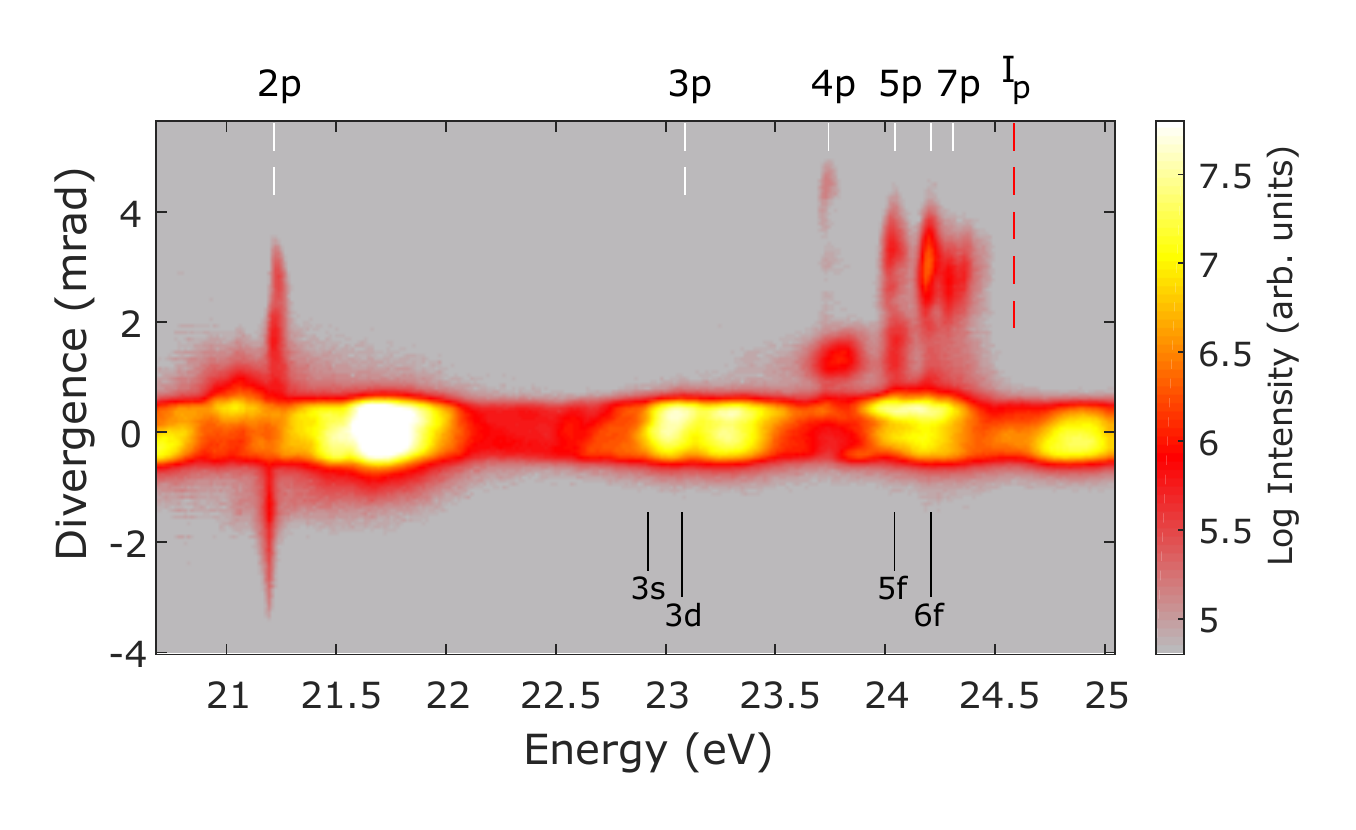}
\caption{Example of XUV spatial control using OOM from a manifold of excited $np$ states in helium. The unperturbed $2p$-$7p$ energies are indicated in white, and the ionization energy in red. The $2p$ emission is redirected both up and down by the 800\,-\,nm IR pulse, whereas the high-lying $np$ emission is only redirected up. States pertinent to later discussions are shown in black. The state energy levels are taken from Ref. \cite{martin_Improved_1987}.
}
\label{fig:Fig1}
\end{figure}

In this article we demonstrate that the OOM technique can be used to probe unknown, nonlinear Stark phases.
In particular, we reveal the intensity dependence of the Stark phase for the low-lying $1s2p$ state in helium (hereafter we omit the passive $1s$ occupation label). We coherently excite the manifold of higher-energy $np$ Rydberg states as a reference and observe that the spatial redirection of the XUV light from the $2p$ transition is different relative to the higher-lying $np$ states.
Significantly, we find that the $2p$ energy shift changes sign as a function of intensity, so that if the $2p$ emission is redirected down at low intensity, it will be redirected up at high intensity. In practice we observe $2p$ emission in both directions at higher peak intensities, because both high and low intensity regions of the IR beam contribute to the redirection, as illustrated in Fig.~\ref{fig:Fig1}.
Solutions of the coupled Maxwell wave equation and the time-dependent Schr\"odinger equation (MWE-TDSE) reproduce the main features of the experimental results. They allow us to understand the observed $2p$ state behavior in terms of a transition from a regime of strong near-resonant coupling with nearby states at low intensity, to a regime of nonresonant free-electron-like behavior at high intensity.

\section{PRINCIPLE}
\begin{figure}[ht]
\includegraphics[trim = {0 0 0 0}, width=\linewidth]{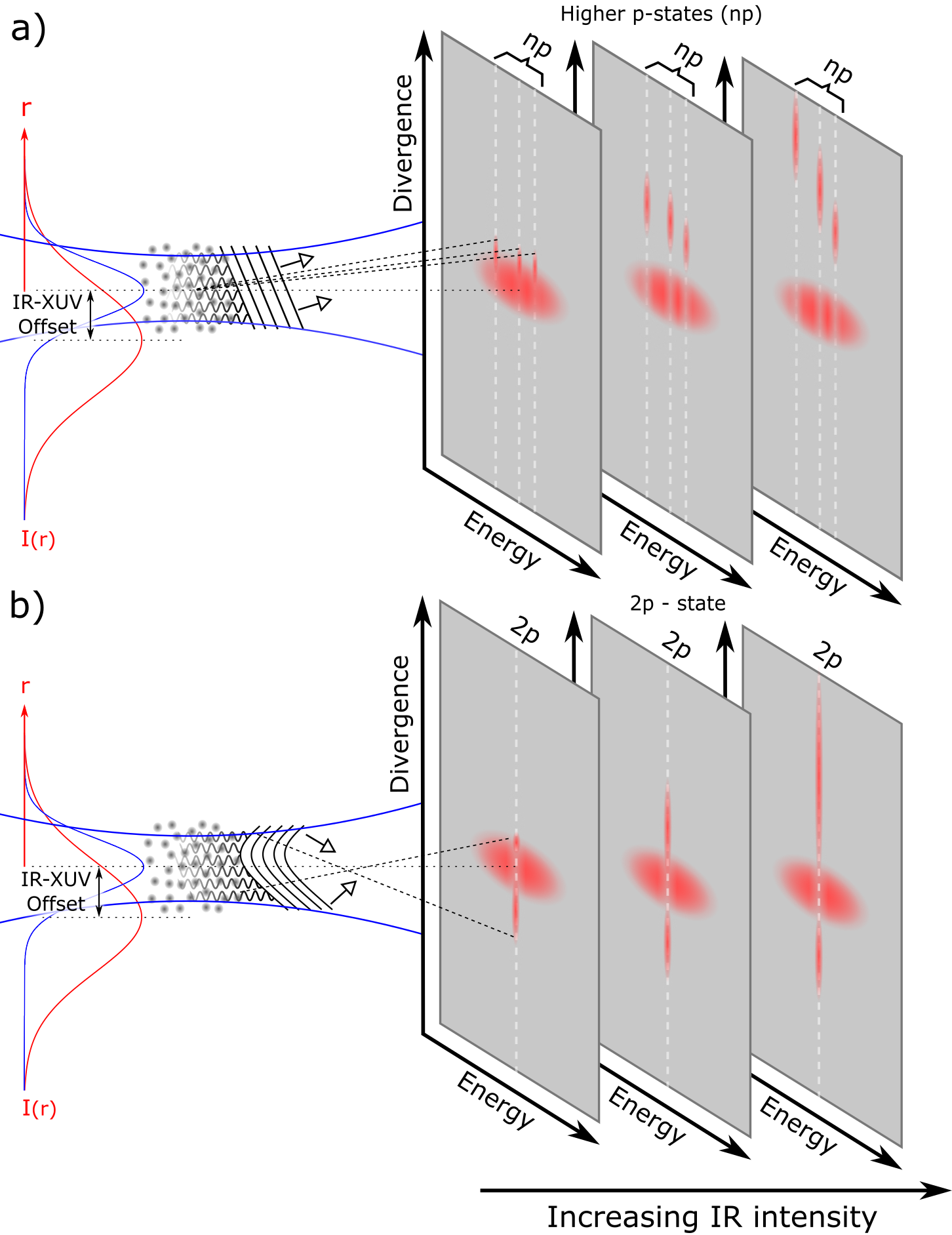}
\caption{Illustration of OOM redirection for (a) linear and (b) nonlinear Stark phase behavior. 
A small (blue in color version) pump XUV beam excites the atoms. (a) Following interaction with a spatially offset, larger (red in color version) probe IR beam, the XUV emission phase front can become tilted if the Stark phase response is approximately linear, as for the $np$ states. (b) A nonlinear Stark response can result in the phase front being tilted in one direction at low intensity, and the other direction at high intensity. The amount of phase accumulation, and consequently the phase gradient spatial profile, is determined by the IR intensity and spatial distribution across the XUV pump focus.
}
\label{fig:Fig2}
\end{figure}

The principle of OOM \cite{bengtsson_spacetime_2017,bengtsson_ultrafast_2019} and how it may be used to probe the intensity dependence of the Stark phase is illustrated in Fig.~\ref{fig:Fig2}. A broadband, coherent XUV pump pulse excites a time-dependent dipole moment, which leads to coherent emission in the forward direction at a number of resonant frequencies  \cite{brewer_optical_1972,hopf_theory_1973}. The long lifetime of the resonances is reflected as sharp absorption features in the spectral domain. An IR probe pulse following the XUV excitation interacts with the target and produces a spatial phase gradient through the intensity-dependent Stark phase, thereby modifying the XUV wavefront and redirecting the emission. 
This happens because the phase gradient 
yields a transverse contribution to the wave vector, $k_{\perp} = d\phi_s/dr$, where $\phi_s$ is the accumulated Stark phase, which alters the direction of wave-vector phase matching.
Since the ac Stark shift is state specific, the emission associated with different excited states can be redirected in different ways by the IR interaction.

To understand the expected behavior of the OOM redirection we consider the spatial dependence of the accumulated Stark phase in the limit where ionization can be ignored \cite{bengtsson_spacetime_2017}:
\begin{equation}
\phi_s({ \bf r}) =\frac{1}{\hbar} 
\int_{\tau_{\hspace*{-0.125em}\rm \textsc{\tiny IR}}}\Delta E({\bf r},t)\,dt,
\label{eq:eq1}
\end{equation}
\noindent where $\Delta E({\bf r},t)$ is the intensity-dependent Stark shift of a specific resonance, $\hbar$ is the reduced Planck constant, and $\origtau_{\textsc{\tiny IR}}$ is the total duration of the IR probe pulse. 
For Rydberg states the shift in energy with increasing field intensity is positive and close to linear. Spatially offsetting a smaller pump beam and a larger probe beam imprints an approximately linear phase gradient across the pump beam so that all the $np$ emission is redirected in the same direction, as observed in Refs. \cite{bengtsson_spacetime_2017,bengtsson_ultrafast_2019} [upward in Fig. \ref{fig:Fig2}(a)]. 
If, however, the intensity-dependent phase shift for a state as a function of intensity is nonlinear, as in Fig.\,\ref{fig:Fig2}(b), the phase front of the emission can be altered in a more complex way.
In particular, if the Stark phase decreases for low intensity and increases at high intensity, the XUV emission can be redirected through both negative and positive divergence angles, resulting in an effective beamsplitter for XUV light. 
 
With the pump and probe beams offset as in Fig.\,\ref{fig:Fig2}, the redirected light can be adjusted via the focal overlap between the pump and probe beams, and the spatial intensity profile of the probe pulse at the target.
For resonances long lived with respect to the duration of the pulses, redirection can occur many tens or hundreds of femtoseconds after the excitation pulse has passed, allowing this measurement to be performed outside of temporal overlap of the pump and probe pulses. For the OOM technique, the lifetime of the excited state must be sufficient for an appreciable Stark-shifting to occur, enabling redirection. Redirection from short-lived states could require shorter pulse durations to satisfy the condition for IR-free XUV excitation of the excited ensemble.

\section{EXPERIMENTAL SETUP}
The experimental setup is a pump-probe scheme where
both pulses are derived from the same  1\,-\,kHz repetition rate, 800\,nm titanium-sapphire laser system producing pulses of $\sim$20\,fs duration. Annular mirrors are used to spatially separate and recombine the pump and probe beam-paths. The outer, annular part of the IR beam is focused into a pulsed gas jet of argon atoms to produce the pump XUV light through high-order harmonic generation (HHG) \cite{mcpherson_studies_1987, ferray_multiple-harmonic_1988, schafer_above_1993, corkum_plasma_1993, lewenstein_theory_1994}. To shift the 13th harmonic into resonance with the $1s$-$2p$ transition in the helium target gas, the HHG process is driven at sufficient intensity to induce blueshifting of the generated harmonics \cite{wood_measurement_1991}. This blueshifting, along with overlaid second-order diffraction components from the diffraction grating, produces the observed near continuous harmonic spectrum detected on axis (Fig.\,\ref{fig:Fig1}). An iris is positioned downstream in the HHG beam path to limit the divergence of the XUV beam and thereby suppress any off-axis emission in the far field that is not due to the IR interaction. This iris also acts to reduce the residual fundamental light from the HHG process. The inner part of the IR beam bypasses the HHG gas and serves as the probe. Both pump and probe beams are focused into the target helium gas using a toroidal mirror. 
Through imaging we measure the probe focus to be $\sim$160\,$\mu$m full width at half maximum (FWHM). From the ability to redirect the XUV $np$ emission either up (as in Fig.\,\ref{fig:Fig1}) or down by adjusting the XUV-IR spatial offset, we deduce that the XUV focus is smaller than this.
The beams are recombined at a small angle, and the probe is offset spatially from the pump in the interaction region to capture the steepest slope of the IR spatial intensity distribution. The delay between the pump and probe pulses is controlled using a precision translation stage, and the delay of the IR probe used in the following measurements is several tens of femtoseconds after temporal overlap.
The helium pressure has been adjusted to optimize the $2p$ emission and avoid effects of resonant pulse propagation (RPP) \cite{vanBurck_Coherent_1999,liao_beyond_2015}. The spectrally resolved spatial profile of the XUV light is recorded in the far field using a flat-field spectrometer, with a microchannel plate detector, imaged by a CCD camera. The probe intensity in the interaction region is controlled using a motorized, zero-aperture iris
after the focus in the IR beam path.

\section{RESULTS}
\subsection{Experiment}
\begin{figure}
\includegraphics[width=\linewidth]{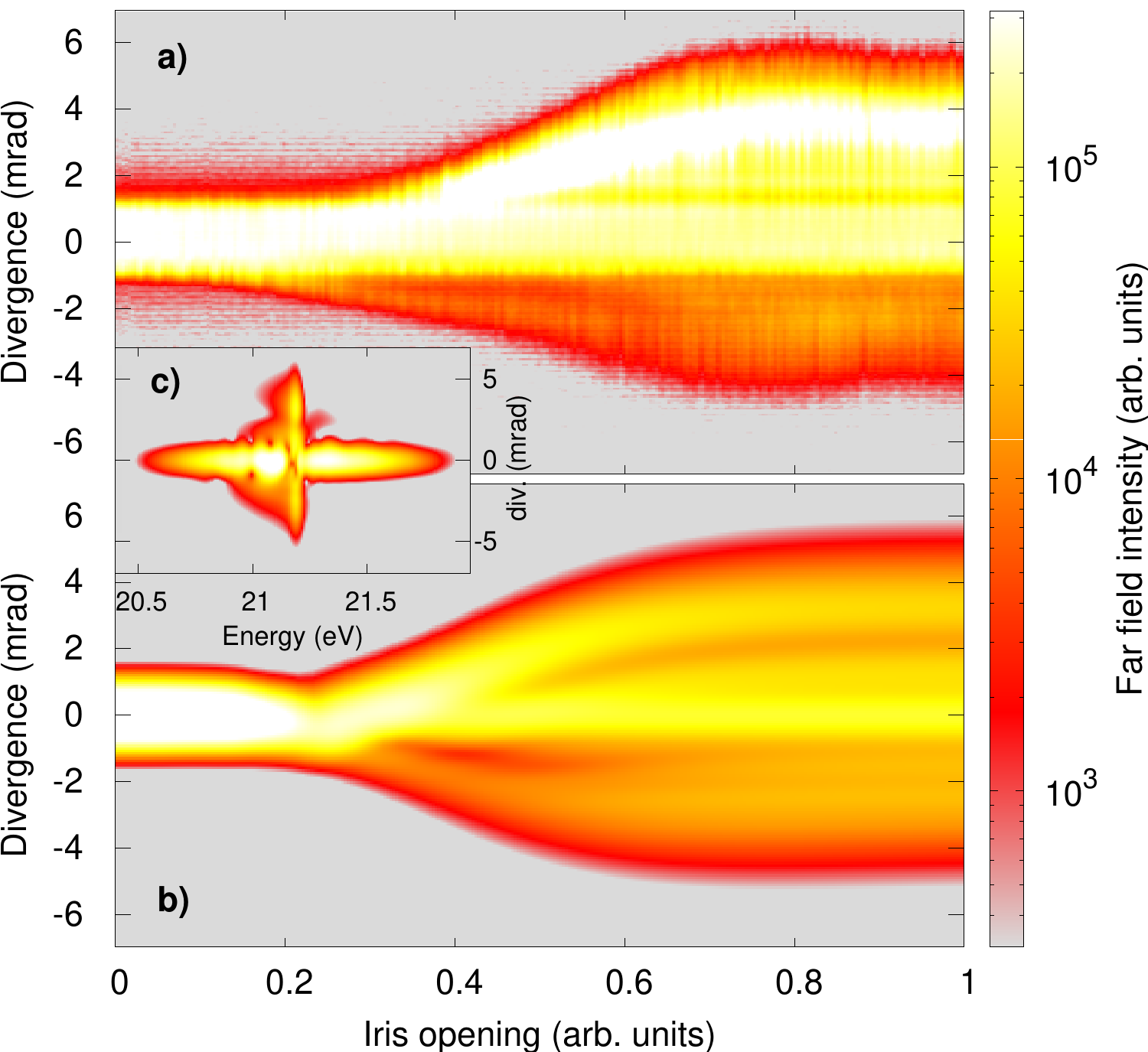}
\caption{
Far-field divergence of the $2p$ emission in (a) the experiment and (b) the calculation. 
(c) The calculated spatiospectral profile of the $2p$ emission for a fully open iris.
}
\label{fig:Fig4}
\end{figure}

 Figure ~\ref{fig:Fig4}(a) shows the evolution of the XUV emission from a narrow energy region around the $2p$-state excitation energy with an iris-opening parameter that varies from 0 (fully closed) to 1 (fully open). Note that the exact mapping between this opening parameter and the actual iris diameter is not perfectly known. The estimated IR peak intensity for the fully open iris is $9\times10^{12}$~W/cm$^2$. The effect of the iris is twofold since it changes both the total energy in the probe beam and its confocal parameter. 
The figure shows that at low intensity (up to iris opening $\approx 0.35$), the $2p$ emission is redirected only downward (opposite to the $np$ emission), whereas  at higher intensities it splits and is redirected both up and down. This indicates that the intensity dependence of the accumulated Stark phase changes sign, or, equivalently, that the shift in energy changes from being negative to being positive.  

\subsection{Theory}
For comparison with experiment, Fig.~\ref{fig:Fig4}(b) shows the  $2p$ emission calculated by solving the coupled MWE-TDSE for a He gas interacting with two spatially offset XUV and IR fields \cite{gaarde_transient_2011}. The $2p$-resonant XUV pump pulse duration is 4\,fs, with a focus of 28\,$\mu$m FWHM and a peak intensity of $1\times10^{11}$~W/cm$^2$, and the 800\,nm probe pulse duration is 27\,fs, with a focus of 56\,$\mu$m FWHM and a peak intensity of $1\times10^{13}$~W/cm$^2$ when the iris is fully opened. The two pulses are delayed with respect to each other by 40\,fs and spatially offset by 35\,$\mu$m. We use a thin 10 $\mu$m He gas medium with a density of $5\times10^{18}$~cm$^{-3}$ to avoid the effects of RPP. To account for the noncylindrical symmetry, the MWE calculations were performed in one transverse direction (1D). This means that the iris in the calculations, which is applied before focusing the IR beam, does not exactly replicate the effect of the experimental iris on the two-dimensional (2D) beam. In particular, the intensity of the 1D beam increases too slowly as the 1D iris diameter is increased as compared to the experiment. To compensate for this, we multiply the intensity after the aperture, $I_a$, by the square of the intensity loss, $I_a/I_0$, where $I_0$ is the intensity before the aperture. The two factors of $I_a/I_0$ mimic the extra drop in intensity due to the energy loss and the increased confocal parameter. 

The calculations can also provide further insight into the $2p$ emission. Figure~\ref{fig:Fig4}(c) shows the calculated far-field, spatiospectral profile of the XUV light near the $2p$ state for a fully open iris, clearly exhibiting both up- and down-directed emission. In the calculation, we can block out selected parts of the near field interaction region, which alters the far-field signal. From this we confirm the picture illustrated in Fig.\,~\ref{fig:Fig2}(b): the downward $2p$ emission comes from the upper part of the probe beam where the intensity is low, and the upward 2p emission comes from the lower part of the probe beam where the intensity is high.
Figure~\ref{fig:Fig4}(b) shows the calculated behavior as a function of iris opening. Allowing for the differences between the experiment and theory discussed above,
the general features of the calculated behavior agree very well with those of the experiment,
both in terms of the down-only redirection at low intensity, and the asymmetry between up and down-directed emission at higher intensities. From the calculations we find that the detailed behavior as a function of iris opening, especially in terms of the asymmetry between the up and down emission, is sensitive to the peak IR intensity, the relative sizes of the pump and probe beams, and in particular to the spatial offset of the pump-probe foci. This suggests that more precise experimental control over the probe spatial profile, for example through the use of spatial light modulators \cite{hasegawa_massively_2016,sun_four-dimensional_2018}, could allow for future reconstruction of the intensity dependence of state-resolved Stark shifts from the experimental results, and to finely control and tailor the XUV emission in space and time.    

\begin{figure}
\includegraphics[width=\linewidth]
{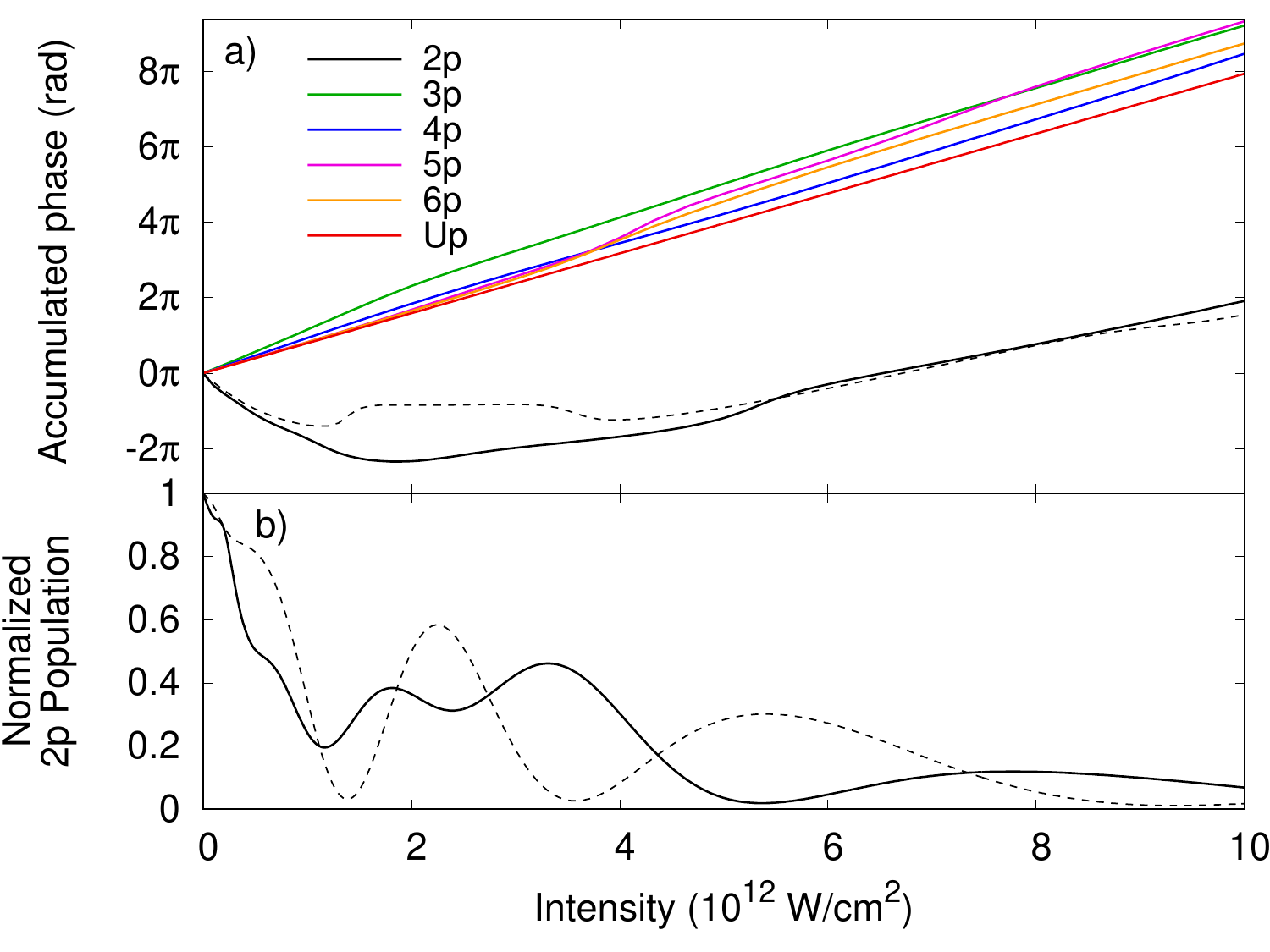}
\caption{(a) TDSE calculation of the total IR-induced phase accumulation for the different $np$ states in helium after interaction with a resonant 4\,-\,fs, $1\times10^{11}$~W/cm$^2$ pump and a subsequent 40\,-\,fs-delayed, 800\,-\,nm, 27\,-\,fs IR probe pulse for a range of different probe peak intensities. The lower solid line corresponds to the $2p$ state, while the upper solid lines correspond to the higher $p$ states, and $U_\mathrm{p}$. (b) $2p$ population at the end of the IR pulse normalized by the $2p$ population at the end of the XUV pulse. The dashed lines correspond to the $2p$ (a) phase and (b) population in the same conditions but with an 829\,-\,nm IR pulse which drives near-resonant two-photon Rabi oscillations between the $2p$ and the $5f$ states. Rapid phase variations are observed at intensities matching near zeros in the $2p$ state populations.}
\label{fig:Fig3}
\end{figure}

Finally, to understand the observed intensity dependence of the $2p$-emission redirection, Fig.~\ref{fig:Fig3}(a) shows the TDSE-calculated accumulated Stark phase for each of the excited states discussed in this paper. The intensity axis denotes the peak intensity of the same 800\,-\,nm, 27\,-\,fs IR pulse used in Fig.~\ref{fig:Fig4}(b), and the phase is extracted at the end of the IR pulse by projecting onto the field free states. The accumulated phase due to a Stark shift equal to the ponderomotive energy
$\Delta E = U_\mathrm{p}$
is shown for comparison and marks the simplest possible linear Stark phase. This figure shows that the accumulated phase increases approximately as $U_\mathrm{p}\tau_{\rm IR}$ for the 3p and higher-lying $np$ states (upper solid lines). The phase of the $2p$ state (lower solid line), however, exhibits a completely different behavior. It drops rapidly at low intensity, below approximately $1.9\times 10^{12}$~W/cm$^2$, then reverses and increases almost linearly at higher intensity, although slower than the higher $np$ states. These general trends are in good agreement with the results discussed above, and can be understood with the following considerations. At low intensity, the $2p$ state couples strongly to the $3s$ and $3d$ states, which are in close to one-photon resonance with it. Indeed, we find that the low-intensity behavior of the $2p$ phase can be accurately reproduced with a three-level model using only the $2p$, $3s$ and $3d$ states (not shown). We also find, as expected for near-resonant interactions, that the sign of the $2p$-$3s$ and $2p$-$3d$ detuning controls the sign of the low-intensity phase shift. The $2s$ state, which is below the $2p$ state by about half an IR photon, is too far detuned to play a significant role. 
Conversely, at high intensities, the electric field strongly distorts the potential felt by the electron so that it behaves increasingly like a free electron in an oscillating field, and the $2p$ state presents a near-linear phase more similar to the higher-lying $np$ states.

At low and moderate intensities, the IR field also enables near-resonant two-photon coupling between the $2p$ and higher-lying $nf$ states \cite{fushitani_femtosecond_2016} that drives Rabi oscillations between these states, as can be seen in the $2p$ population shown in Fig.~\ref{fig:Fig3}(b). These oscillations are highly sensitive to the IR wavelength and are best observed at a slightly longer  wavelength (829\,nm) than the one used in the experiment. The longer-wavelength $2p$ population and phase are shown as dashed lines in Figs.~\ref{fig:Fig3}(a) and (b). Note that the minima in the $2p$ population are associated with rapid variations of the phase (near  $1.5\times10^{12}$ and $3.5\times10^{12}$~W/cm$^2$), as expected for Rabi flopping \cite{wu2013ultrafast}. 
This provides another interesting perspective on XUV spatial control through OOM: in the resonant case, both the phase and the amplitude of the XUV field can be modulated through IR control of the Stark shift and the population of the resonant state. 

\section{SUMMARY}
In summary, we have used the all-optical OOM technique to 
probe the Stark-induced phase change of excited states in matter.
We have experimentally observed the change of sign of the $2p$-state phase accumulation as the intensity of the nonresonant IR field is increased, in good agreement with MWE-TDSE based calculations. This result opens the possibility for the future study of Stark phases in more complicated atoms or molecules, where the states and/or their dipole couplings may be less well known,
and could even allow for reconstruction of the phase accumulation from the experimental result given tighter control over the experimental parameters.  We also emphasize the potential for the OOM technique to be used to probe unknown Stark phases of states embedded in the continuum, which, although beyond the scope of the work presented here, would be interesting to study in future experiments.
This work also highlights the potential for the OOM technique to control XUV frequency light in different ways, such as by creating variable beam splitters in the XUV by exploiting the nonlinear response of states to IR intensity changes. 

\section*{ACKNOWLEDGMENTS}
This research was partially supported by the Swedish Foundation for Strategic Research, the
Crafoord Foundation, the Swedish Research Council, the Wallenberg Centre for Quantum Technology (WACQT) funded by the Knut and Alice Wallenberg Foundation, and the Royal Physiographic Society of Lund. Research at Louisiana State University was supported by the US Department of Energy, Office of Science, Basic Energy Sciences, under Contract No. DE-SC0010431. Portions of this research were conducted with high performance
computing resources provided by Louisiana State University (http://www.hpc.lsu.edu) and Louisiana Optical Network Infrastructure (http://hpc.loni.org).

\bibliography{Bibliography}

\end{document}